\documentstyle[prd,aps,preprint,tighten,epsfig]{revtex}

\begin{document}
\hbadness=10000 \pagenumbering{arabic}

\title{Branching ratios of $B^+ \to D^{(*)+}K^{(*)0}$ decays in perturbative QCD approach}

\author{Ge-Liang Song \footnote{Email: songgl@mail.ihep.ac.cn}, Cai-Dian  L\"{u} \footnote{Email: lucd@mail.ihep.ac.cn}}
\address{\small CCAST (World Laboratory), P.O. Box 8730, Beijing 100080,
China; \\
\small  Institute of High Energy Physics, CAS, P.O. Box 918(4),
Beijing 100039, China}
 \maketitle

\date{}


\begin{abstract}
We study the rare decays $B^+ \to D^{(*)+}K^{(*)0}$, which can
occur only via annihilation type diagrams in the standard model.
We calculate all of the four modes, $B \to PP,~ VP,~ PV,~ VV$, in
the framework of perturbative QCD approach and give the branching
ratios of the order about $10^{-6}$.
\end{abstract}
%

\section{Introduction}
More and more data of $B$ decays are being collected at the two
$B$ factories, Belle and BaBar. The original approach to
non-leptonic $B$ decays based on the factorization approach (FA)
\cite{1wsb}, which succeeded in calculating the branching ratios
of many decays \cite{2aklch}. FA is a simple method, by which
non-factorizable and annihilation contributions are neglected.
Although calculations are easy in FA, it suffers the problems of
scale, infrared-cutoff and gauge dependence \cite{3cly}.
Especially it is difficult to explain some observed branching
ratios of $B$ decays, such as $B \to J/\psi K^{(*)}$ \cite{4yl}.
To improve the theoretical application \cite{5bbns} and understand
why the simple FA works so well \cite{6kls,7luy}, some methods
have been brought forward and developed. One of them is the
perturbative QCD (PQCD) approach developed by Brodsky and Lepage
\cite{8lb79,9lb80}, under which we can calculate the annihilation
diagrams as well as the factorizable and non-factorizable
diagrams.

It is consistent to calculate branching ratios of $B$ decays in
PQCD approach, as we will explain its framework in the next
section. It has been applied in the non-leptonic $B$ decays
\cite{6kls,7luy,10chang,11PQCD} successfully. In the case of $B^+
\to D^{(*)+}K^{(*)0}$ decays, which is a kind of pure annihilation
type decays, the physics picture of PQCD is as follows, shown in
Fig. 1. A $W$ boson exchange causes $\bar{b}u \to \bar{s}c $, and
the $\bar{d}d$ quarks are produced from a gluon. In the rest frame
of the $B$ meson, the $d$ and $\bar{d}$ quarks in the
$D^{(*)+}K^{(*)0}$ mesons each has momentum $\cal{O}$$(M_B/4)$, so
the gluon producing them has $q^2 \sim \cal{O}$$(M_B^2/4)$. It is
a hard gluon according to the mass of $B$ meson. So we can
perturbatively treat these decays and use PQCD approach like other
pure annihilation type $B$ decays \cite{112} .

In the next section, we explain the framework of PQCD briefly. In
section 3, we give the analytic formulae for the decays $B^+ \to
D^{(*)+}K^{(*)0}$. In section 4, we show the numerical results and
theoretical errors of the four modes respectively. Finally, we
draw a conclusion in section 5.

\section{Framework}
The PQCD approach divides the process into hard components, which
are treated by perturbative theory, and non-perturbative
components, which are put into the hadron wave function. The
hadron wave function can be extracted from experimental data or
calculated by QCD sum rules method. The decay amplitude can be
conceptually written as the convolution
\begin{equation}
 \mbox{Amplitude}
\sim \int\!\! d^4k_1 d^4k_2 d^4k_3\ \mathrm{Tr} \bigl[ C(t)
\Phi_B(k_1) \Phi_{D_s^*}(k_2) \Phi_K(k_3) H(k_1,k_2,k_3, t)
\bigr], \label{eq:convolution1}
\end{equation}
where $k_i$'s are momenta of light quarks included in each meson,
and $\mathrm{Tr}$ denotes the trace over Dirac and color indices.
The hard components comprise hard part ($H$) and harder dynamics
($C$). $H(t)$ describes the four quark operator and the spectator
quark connected by a hard gluon. It can be perturbatively
calculated, since it includes the hard dynamics characterized by
the scale $t$, where $t\sim\mathcal{O}$ $(M_B/2)$ for $B^+ \to
D^{(*)+}K^{(*)0}$ decays, and the hard gluon's $q^2$ is of the
order $t^2$. $C(t)$ is the Wilson coefficient which results from
the radiative corrections at short distance. In the above
convolution, $C(t)$ includes the harder dynamics at a larger scale
than the $M_B$ scale and describes the evolution of local
four-Fermi operators from $M_W$ down to the scale $t$. The wave
function $\Phi_M$ denotes the non-perturbative components, which
is independent of the specific processes and removes the infrared
cut off dependence in PQCD approach.

According to the conservation of four-momentum, we can obtain
$D^{(*)+}$ and $K^{(*)0}$ meson's energy and momenta in the rest
frame of $B$ meson,
\begin{eqnarray}
      & &E_1=M_B, ~~\left|{\bf P}_1\right|=0,\nonumber\\
      & &E_2=\frac{M_B}{2}(1+r_2^2-r_3^2),~~ E_3=\frac{M_B}{2}(1-r_2^2+r_3^2),\nonumber\\
      & &\left|{\bf P}_2\right|=\left|{\bf P}_3\right|=\frac{M_B}{2}\sqrt{1 - 2\, {{r_2}}^2 +{{r_2}}^4 - 2\,{{r_3}}^2 -
     2\,{{r_2}}^2\,{{r_3}}^2 +{{r_3}}^4},      \label{eq:smoment}
\end{eqnarray}
where the subscript (1,2,3) denote $B^+$, $D^{(*)+}$ and
$K^{(*)0}$ meson respectively, and $r_2= M_{D^{(*)+}}/M_B$, $r_3=
M_{K^{(*)0}}/M_B$. It is convenient to assume that the $D^{*+}$
($K^{*0}$) meson moves in the plus (minus) $z$ direction carrying
the momentum $P_2$ ($P_3$). The longitudinal polarization vectors
of the $D^{*+}$ and $K^{*0}$ are given as
\begin{eqnarray}
 & & \epsilon_{2L}= \frac{1}{2 r_2}
(\sqrt{
     {\left( 1 -
         {{r_3}}^2
         \right) }^2 -
     2\,{{r_2}}^2\,
      \left( 1 +
        {{r_3}}^2
        \right) +{{r_2}}^4 },0,0,1 + {{r_2}}^2 -
  {{r_3}}^2),  \nonumber\\
  & & \epsilon_{3L}= \frac{1}{2
r_3} ({\sqrt{
     {\left( 1-
         {{r_3}}^2
         \right) }^2 -
     2\,{{r_2}}^2\,
      \left( 1 +
        {{r_3}}^2
        \right) +{{r_2}}^4 }},0,0,-1 + {{r_2}}^2 -
  {{r_3}}^2), \label{eq:spolarization}
\end{eqnarray}
which satisfy the normalization
$\epsilon_{2L}^2=\epsilon_{3L}^2=-1$ and the orthogonality
$\epsilon_{2L}\cdot P_2=\epsilon_{3L}\cdot P_3=0$. For simplicity,
we use the light-cone coordinate\footnote{We use the light-cone
coordinate in the convention, where $p^\pm = \frac{1}{\sqrt{2}}
(p^0 \pm p^3)$ and ${\bf p}_T = (p^1, p^2)$.} to describe the
meson's momenta in the rest frame of $B$ meson. After deducing the
analytic formulae of amplitudes, we ignore the terms proportional
to $r_2^2 \sim 0.15$ or $r_3^2 \sim 0.04$\footnote{This
approximation is also adapted in deriving meson wave functions. So
it is consistent to take
eqs.(\ref{eq:lmoment},\ref{eq:lpolarization}). Moreover we ignore
the mass of light pseudo-scalar meson $K^0$.}. Equivalently we
ignore the terms proportional to $r^2$ in $B^+$, $D^{(*)+}$ and
$K^{(*)0}$ meson's momenta and longitudinal polarization vectors.
Therefore eqs.(\ref{eq:smoment},\ref{eq:spolarization}) in the
light-cone coordinate are corresponding to
\begin{equation}
       P_1 = \frac{M_B}{\sqrt{2}} (1,1,{\bf 0}_T), \  \  P_2 =
       \frac{M_B}{\sqrt{2}} (1-r_3^2,r_2^2,{\bf 0}_T), \  \  P_3 =
       \frac{M_B}{\sqrt{2}} (r_3^2,1-r_2^2,{\bf 0}_T), \label{eq:lmoment}
\end{equation}
\begin{equation}
  \epsilon_{2L}= \frac{1}{\sqrt{2} r_2}
(1-r_3^2,-r_2^2,{\bf 0}_T),\ \   \epsilon_{3L}= \frac{1}{\sqrt{2}
r_3} (-r_3^2,1-r_2^2,{\bf 0}_T), \label{eq:lpolarization}
\end{equation}
respectively. The transverse polarization vectors can be adapted
directly as $ \epsilon_{2T}= (0,0,{\bf 1}_T) $, $ \epsilon_{3T}=
(0,0,{\bf 1}_T) $. We denote the light (anti-)quark momenta in
$B^+$, $D^{(*)+}$ and $K^{(*)0}$ mesons as $k_1 = (x_1
P_1^+,0,{\bf k}_{1T})$, $k_2 = (x_2 P_2^+,0,{\bf k}_{2T})$, and
$k_3 = (0, x_3 P_3^-,{\bf k}_{3T})$ respectively. Integrating eq.
(\ref{eq:convolution1}) over $k_1^-$, $k_2^-$, and $k_3^+$, we
obtain
\begin{eqnarray}
 \mbox{Amplitude}
\sim \int\!\! d x_1 d x_2 d x_3
b_1 d b_1 b_2 d b_2 b_3 d b_3 \\
\mathrm{Tr} \bigl[ C(t) \Phi_B(x_1,b_1) \Phi_{D_s}(x_2,b_2)
\Phi_K(x_3, b_3) H(x_i, b_i, t) S_t(x_i)\, e^{-S(t)} \bigr],
\label{eq:convolution2}
\end{eqnarray}
where $b_i$ is the conjugate space coordinate of $k_{iT}$, and $t$
is the largest energy scale in $H$, as a function in terms of
$x_i$ and $b_i$. The large logarithms $\ln( M_W/t)$ resulting from
QCD radiative corrections to four quark operators are absorbed
into the Wilson coefficients $C(t)$. The inclusion of $k_T$ brings
in one kind of large logarithms $\ln^2(Pb)$ from the overlap of
collinear and soft gluon corrections, $P$ denoting the dominant
light-cone component of meson momentum. The other kind of large
logarithms $\ln(tb)$ derives from the renormalization of the
ultraviolet divergences. These two kinds of large logarithms are
summed and lead to Sudakov form factor, $e^{-S(t)}$. It suppresses
the soft dynamics effectively \cite{12soft}. The large double
logarithms $\ln^2 x_i$ are summed by the threshold resummation
\cite{13sum}, and they lead to $S_t(x_i)$ which smears the
end-point singularities on $x_i$. From the brief analysis above,
it can be seen that PQCD is a consistent approach.

\section{Analytic formulae}
\subsection{The wave functions}
We use the wave functions $\Phi_{M,\alpha\beta}$ decomposed in
terms of spin structure. The coming $B$ meson and outgoing
$D^{(*)+}$, $K^{(*)0}$ are as follows:
\begin{equation}
 \Phi_{B}(x,b) = \frac{i}{\sqrt{2N_c}}
\left[ (\not \! P_1 \gamma_5) + M_B \gamma_{5} \right]
\phi_B(x,b),
\end{equation}

\begin{equation}
 \Phi_{D}(x,b) = \frac{i}{\sqrt{2N_c}}
\left[  \gamma_5 (\not \! P_2+ M_{D}) \right] \phi_{D}(x,b),
\end{equation}

\begin{equation}
 \Phi_{D^{*}}(x,b) = \frac{i}{\sqrt{2N_c}}
\left[  \not \! \epsilon (\not \! P_2+ M_{D^*}) \right]
\phi_{D^*}(x,b),
\end{equation}

\begin{eqnarray}
\Phi_{K}(x,b) = \frac{i}{\sqrt{2N_c}} \Bigl[ \gamma_5 \not \! P_3
\phi_K^A(x,b) + M_{0K} \gamma_5 \phi_K^P(x,b) \nonumber \\
 + M_{0K} \gamma_5 (\not v \not n - 1)\phi_K^T(x,b)
\Bigr],
\end{eqnarray}
\begin{eqnarray}
 \Phi_{K^*L}(x,b) = \frac{i}{\sqrt{2N_c}} \Bigl[ M_{K^*} \not \!
 \epsilon_{3L}
\phi_{K^*}(x,b) +\not \! \epsilon_{3L} \not \! P_3
\phi_{K^*}^t(x,b) \nonumber \\ + M_{K^*} I \phi_{K^*}^s(x,b)
\Bigr],
\end{eqnarray}
\begin{eqnarray}
\Phi_{K^*T}(x,b) = \frac{i}{\sqrt{2N_c}} \Bigl[M_{K^*}\not \!
\epsilon_{3T}\phi_{K^*}^v(x)+ \not \! \epsilon_{3T}\not
P_3\phi_{K^*}^T(x)  \nonumber \\  +\frac{M_{K^*}}{P_3\cdot n}
i\epsilon_{\mu\nu\rho\sigma}\gamma_5\gamma^\mu\epsilon_{3T}^\nu
P_3^\rho n^\sigma \phi_{K^*}^a(x)\Bigr],
\end{eqnarray}
where $N_c = 3$ is color's degree of freedom, and $M_{0K} =
M_K^2/(m_u + m_s)$, $v = (0,1,{\bf 0}_T ) \propto P_3$, $n =
(1,0,{\bf 0}_T)$, $\epsilon^{0123} = 1$. The subscripts $L$ and
$T$ denote the wave functions corresponding to the longitudinally
and transversely polarized $K^*$ mesons.
\subsection{The effective Hamiltonian}
The effective Hamiltonian for decay $B^+ \to D^{(*)+}K^{(*)0}$ at
a scale lower than $M_W$ is given by \cite{14buras}
\begin{eqnarray}
 H_{\mathrm{eff}} = \frac{G_F}{\sqrt{2}} V_{ub}^*V_{cs} \left[
C_1(\mu) O_1(\mu) + C_2(\mu) O_2(\mu) \right], \\
  O_1 = (\bar{b}s)_{V-A} (\bar{c}u)_{V-A}, \quad
 O_2 = (\bar{b}u)_{V-A} (\bar{c}s)_{V-A},
\end{eqnarray}
where $C_{1,2}(\mu)$ are Wilson coefficients at renormalization
scale $\mu$, and summation in $\mathrm{SU}(3)_c$ color's index
$\alpha$ and chiral projection, $\sum_\alpha \bar{q}_\alpha
\gamma^\nu(1-\gamma_5)q'_\alpha$ are abbreviated to
$(\bar{q}q')_{V-A}$. The lowest order diagrams of $B^0 \to
D_s^{*-} K^+$ are drawn in Fig. 1. We will choose
$|V_{cs}|=0.996\pm 0.013, \  |V_{ub}|= (3.6 \pm 0.7)\times
10^{-3}$ \cite{15pdgv}. There is no CP violation in the decays,
since only one kind of CKM phase appears in the processes.
Therefore the decay width for CP conjugated mode, $B^- \to
D^{(*)-} \bar{K}^{(*)0}$, equals to $B^+ \to D^{(*)+}K^{(*)0}$
respectively.

\subsection{The decay width}
The total decay amplitude for each mode or helicity state of $B^+
\to D^{(*)+}K^{(*)0}$ is written as
\begin{equation}
 A = f_B F + M , \label{eq:neut_amp}
\end{equation}
where $f_B$ is the decay constant of $B$ meson, and the overall
factor is included in the decay width with the kinematics factor.
$F(M)$ stands for the amplitude of (non-) factorizable
annihilation diagrams in Fig. 1a,b (c,d). We exhibit their
explicit expressions and subscripts of $F$ and $M$ according to
the modes and helicity states respectively in the appendix.  The
decay width for each mode of these decays is given as
\begin{equation} \Gamma =\frac{G_F^2
M_B^3}{128\pi} (1-r_2^2)
 (\sum_{\sigma})|V_{ub}^*V_{cs} A_\sigma|^2\;, \label{dr1}
\end{equation}
where the subscript $\sigma$ denotes the helicity states of the
two vector mesons with $L(T(1,2))$ standing for the longitudinal
(transverse) component in the case of $B^+ \to D^{*+}K^{*0}$
decay, as shown in the appendix.

\section{Numerical results}

In the numerical analysis, we adopt the $B$ meson wave function as
\cite{6kls,7luy}
\begin{equation}
\phi_B(x,b) = N_B x^2(1-x)^2 \exp \left[ -\frac{M_B^2\ x^2}{2
\omega_b^2} -\frac{1}{2} (\omega_b b)^2 \right],\label{eq:Bwave}
\end{equation}
with the shape parameter $\omega_b$ and the normalization constant
$N_B$ being related to the decay constant $f_B$ by normalization
\begin{equation}
 \int_0^1 \!\! dx\  \phi_M (x, b=0)
= \frac{f_M}{2\sqrt{2N_c}}, \label{eq:normalization}
\end{equation}
which is also right for $D^{(*)}$ meson, i.e. $M = B,D^{(*)}$.

For $D^{(*)}$ meson wave function, we use two types. The first
kind \cite{16kls} is
\begin{eqnarray}
(I)& &\ \ \ \ \ \ \ \phi_{D^{(*)}}(x) = \frac{3}{\sqrt{2 N_c}}
f_{D^{(*)}} x(1-x)\{ 1 + a_{D^{(*)}} (1 -2x) \}\exp
\left[-\frac{1}{2} (\omega_{D^{(*)}} b)^2 \right],
\end{eqnarray}
in which the last term, $\exp \left[-\frac{1}{2}
(\omega_{D^{(*)}}b)^2 \right]$, derived from the $k_T$
distribution. By taking same parameters, we neglect the difference
between the $D$ and $D^*$ mesons wave functions, since the $c$
quark is much heavier than the $\bar d$ quark, and the mass
difference between two mesons is little. The second kind
\cite{17many} is
\begin{eqnarray}
(II)& &\ \ \ \ \ \ \ \phi_{D^{(*)}}(x) = \frac{3}{\sqrt{2 N_c}}
f_{D^{(*)}} x(1-x)\{ 1 + a_{D^{(*)}} (1 -2x) \},
\end{eqnarray}
which is fitted from the measured $B \to D^{(*)}\ell \nu$ decay
spectrum at large recoil. The absence of the last term like $(I)$
is due to the insufficiency of the experiment data.

The $K^{(*)}$ wave functions \cite{18ps,19vect} we adopt are
calculated by QCD sum rules. To abridge the context, we list them
and the corresponding parameters in the appendix.

The other input parameters are listed below:
\begin{eqnarray}
& & f_B = 190\; {\rm MeV}\;,\;\; f_{D} = 240\; {\rm
MeV}\;,\;\;f_{D^*} = 240\; {\rm MeV}\;,\;\;
\nonumber \\
& & f_K = 160\; {\rm MeV}\;,\;\; f_{K^*} = 200\; {\rm
MeV}\;,\;\;f_{K^*}^T = 160\; {\rm
MeV}\;,\;\;\omega_{D^{(*)}}=0.2\;{\rm GeV},\;\;
\\
& &M_{0K} = 1.60\; {\rm GeV}\;,\;\;\omega_b=0.4\;{\rm GeV}\;,\;\;
a_{D^{(*)}} = 0.3\;,\;\; C_D=0.8\;,\;\; C_{D^*}=0.7\;,
\label{eq:para}
\\
& &M_B = 5.279\; {\rm GeV}\;,\;\; M_b = 4.8\; {\rm GeV}\;,\;\; M_D
= 1.869\; {\rm GeV}\;,\;\; M_{D^*} = 2.010\; {\rm GeV}\;,\;\;
\nonumber\\
 & &M_t=170\; {\rm GeV}\;,\;\;M_W = 80.4\;{\rm GeV}\;,\;\;\tau_{B^\pm}=1.674\times 10^{-12}{\rm s}\;,G_F=1.16639\times
10^{-5}\;{\rm GeV}^{-2}\;, \label{eq:parapdg}
\end{eqnarray}
where the Fermi coupling constant $G_F$, the masses and life times
of particles refer to \cite{20pdg}.

With the analytic formulae and parameters above, we get the
branching ratios of $B^+ \to D^{(*)+}K^{(*)0}$ shown in Table
\ref{tb:1}, \ref{tb:11}, \ref{tb:2}, \ref{tb:22} for two kinds of
$D^{(*)}$ wave functions respectively. The magnitude according to
$D^{(*)}$ wave function $I$ is about 60 percent of the one
corresponding to $D^{(*)}$ wave function $II$. The difference can
tell the correct $D^{(*)}$ wave function by the experiment data in
the future.

\begin{figure}[h]
 \begin{center}
   \center
  \includegraphics[width=14cm]{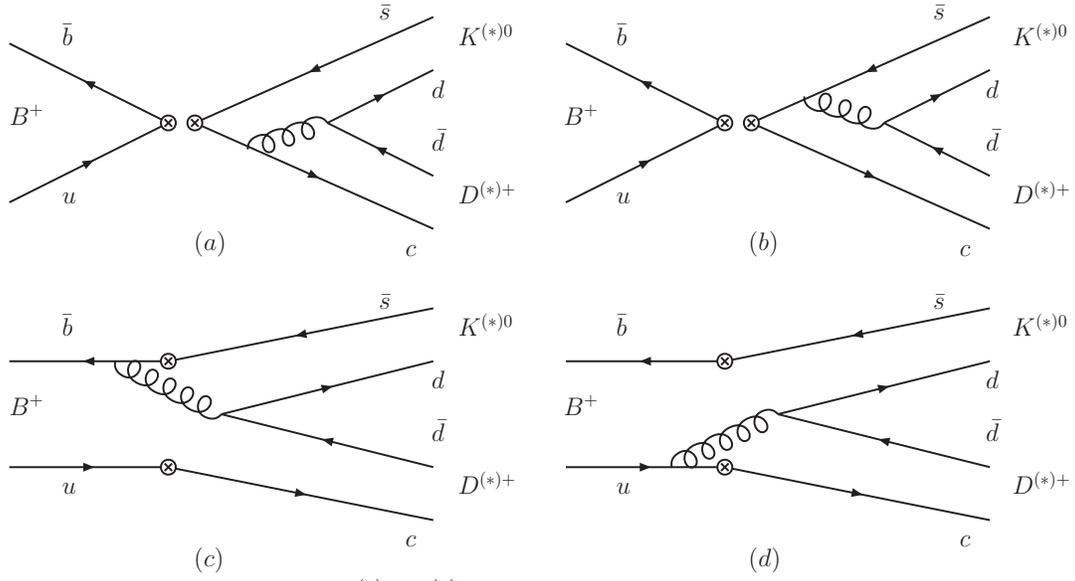}\\
\caption{Diagrams for $B^+ \to D^{(*)+} K^{(*)0}$ decays. The
factorizable diagrams (a), (b) contribute to $F$,
 and nonfactorizable (c), (d) contribute to $M$.}
 \end{center}
 \label{fig1}
\end{figure}

For each mode of $B^+\to D^{(*)+}K^{(*)0}$ decays, the
contribution of the factorizable and non-factorizable annihilation
diagrams is the same order, although $F$ is proportional to Wilson
coefficient $C_2+C_1/3$, which is $\mathcal{O}$ $(1)$, and
non-factorizable annihilation diagram contribution is proportional
to $C_1$, which is about 30 percent of $C_2+C_1/3$. Since the
counteraction influence between Fig. \ref{fig1}a,b of $F$ is
heavier than that between Fig. \ref{fig1}c,d of $M$ by the reason
of the more similar propagators in Fig. \ref{fig1}a,b. The
magnitude comparison can bee seen directly from Table \ref{tb:1},
\ref{tb:11}, \ref{tb:2}, \ref{tb:22}.
\begin{table}[htbp]
\begin{center}
\begin{tabular}[t]{c|ccc}
  $ $ & $B \to DK$ & $B \to D^*K$ & $B \to DK^*$ \\
 \hline \hline
$f_BF$ & $-1.56+1.21i$ & $-1.86+2.20i$ & $1.17-1.65i$  \\
 $M$   & $1.03+1.29i$  & $-1.01-0.78i$ & $-1.75-0.89i$  \\
  \hline
 $A$ & $-0.52+2.50i$ & $-2.87+1.41i$ & $-0.58-2.54i$  \\
   \hline          \hline
$\mathrm{Br} (10^{-6})$ & $0.93$ & $1.42$ & $0.96$ \\
\end{tabular}
\end{center}
\caption{The branching ratios of the three decay modes and
amplitudes $(10^{-2} GeV)$ in terms of the factorizable,
non-factorizable diagrams and the sum of them according to the
$D^{(*)}$ wave function I. }
 \label{tb:1}
\end{table}

\begin{table}[htbp]
\begin{center}
\begin{tabular}[t]{c|ccc}
  $ $ & $(B \to D^*K^*)_{L}$ & $(B \to D^*K^*)_{T1}$ & $(B \to D^*K^*)_{T2}$ \\
 \hline \hline
$f_BF$ & $-0.17-3.44i$ & $1.58-3.47i$ & $-0.37-0.22i$  \\
 $M$   & $1.73+0.35i$  & $-0.09-0.41i$ & $0.009+0.005i$  \\
  \hline
  $A$ & $1.56-3.08i$ & $1.49-3.87i$ & $-0.36-0.21i$  \\
   \hline          \hline
$\mathrm{Br} (10^{-6})$ & $1.67$ & $2.40$ & $0.02$ \\
   \hline
$Total~ \mathrm{Br} (10^{-6})$ &$4.09$ \\
\end{tabular}
\end{center}
\caption{The branching ratios of $B \to D^*K^*$ decay and helicity
amplitudes $(10^{-2} GeV)$ in terms of the factorizable,
non-factorizable diagrams and the sum of them according to the
$D^{*}$ wave function I. }
 \label{tb:11}
\end{table}

\begin{table}[htbp]
\begin{center}
\begin{tabular}[t]{c|ccc}
 $ $ & $B \to DK$ &
$B \to D^*K$ & $B \to DK^*$ \\
 \hline \hline
$f_BF$ & $-2.38+1.56i$ & $-2.47+2.94i$ & $1.50-2.32i$  \\
 $M$ & $1.37+1.49i$ & $-1.24-0.87i$ & $-2.18-1.01i$  \\
  \hline
 \hline
 $A$ & $-1.01+3.05i$ & $-3.71+2.07i$ & $-0.68-3.32i$  \\
   \hline
$\mathrm{Br} (10^{-6})$ & $1.47$ & $2.52$ & $1.64$  \\
\end{tabular}
\end{center}
\caption{The branching ratios of the four decay modes and
amplitudes $(10^{-2} GeV)$ in terms of the factorizable,
non-factorizable diagrams and the sum of them according to the
$D^{(*)}$ wave function II.}
 \label{tb:2}
\end{table}

\begin{table}[htbp]
\begin{center}
\begin{tabular}[t]{c|ccc}
 $ $ & $(B \to D^*K^*)_{L}$ & $(B \to D^*K^*)_{T1}$ & $(B \to D^*K^*)_{T2}$ \\
 \hline \hline
$f_BF$ & $-0.38-4.61i$ & $2.02-4.62i$ & $-0.44-0.22i$  \\
 $M$   & $2.03+0.37i$  & $-0.14-0.49i$ & $0.01+0.006i$  \\
  \hline
  $A$ & $1.65-4.24i$ & $1.88-5.11i$ & $-0.43-0.22i$  \\
   \hline          \hline
$\mathrm{Br} (10^{-6})$ & $2.89$ & $4.12$ & $0.03$ \\
   \hline
$Total~ \mathrm{Br} (10^{-6})$ &$7.04$ \\
\end{tabular}
\end{center}
\caption{The branching ratios of $B \to D^*K^*$ decay and helicity
amplitudes $(10^{-2} GeV)$ in terms of the factorizable,
non-factorizable diagrams and the sum of them according to the
$D^{*}$ wave function II. }
 \label{tb:22}
\end{table}

From Table \ref{tb:11}, \ref{tb:22}, we can see $|A_{T1}|
> |A_L| \gg |A_{T2}|$ in the case of $B\to VV$ mode. There are two
questions worthy of asking. why is $|A_{T2}|$ so little? why are
$|A_{T1}|$ and $|A_L|$ the same order, though $|A_{T1}|$ is
suppressed at least by the term $r^2$ ($r=r_2$ or $r_3 $)?
According to the amplitudes of $F_{T2}$ and $M_{T2}$, the
contribution of the twist 2 wave function $\phi_{K^*}^T$ is
absent, and the coefficients corresponding to the twist 3 wave
functions $\phi_{K^*}^v$ and $\phi_{K^*}^a$ are just opposite and
counteract each other heavily. Therefore the value of $A_{T2}$ is
too little to consider. To answer the second question, we should
note that $r_2$ is not a serious suppression term, especially when
$r_2$ times 2, $2r_2 \simeq 1$, like the term in $F_{T1}$ and
$M_{T1}$. In the case of $F_{T1}$ and $M_{T1}$, all of the signs
of the sub-amplitudes corresponding to the two twist 3 $K^*$ wave
functions are same, and the terms in the front of the twist 2 wave
function $\phi_{K^*}^T$ do not suffer the heavy suppression of
$r_3$. On the other hand, in $F_L$ ($M_L$) the seemly main
contribution of the twist 2 wave function $\phi_{K^*}$ is offset
by the opposite coefficients in Fig. \ref{fig1}a,b (c,d). Moreover
in Fig. \ref{fig1}a the signs of the coefficients corresponding to
the twist 3 wave function $\phi_{K^*}^t$ and $\phi_{K^*}^s$ are
different. For the reasons above, $|A_{T1}| $ and $|A_L|$ are the
same order.

It should be stressed that there is no arbitrary parameter in our
calculation, but we only know the magnitude of each up to a range.
In Table \ref{tb:I},\ref{tb:II} we show the sensitivity of the
branching ratios to $30$\% change of the parameters in eq.
(\ref{eq:para}) according to the two kinds of $D^{(*)}$ wave
functions respectively. Since the $M_{0K}$ and $\omega_b$'s
uncertainty influences the results very much, we will limit them
to a more appropriate extent. According to \cite{18ps},
\begin{equation}
 1.4 \mbox{ GeV} \leq M_{0K} \leq 1.8 \mbox{ GeV},
\label{eq:m0K}
\end{equation}
the branching ratios are
\begin{equation}
 \mathrm{Br}(B^+ \to D^+ K^0) = \left \{
  \begin{array}{ll}
  0.93_{-0.06}^{+0.06} \times 10^{-6} &  I \\
  1.47_{-0.13}^{+0.13} \times 10^{-6} &  II\\
  \end{array}
\right.
\end{equation}
\begin{equation}
 \mathrm{Br}(B^+ \to D^{*+} K^0) = \left \{
  \begin{array}{ll}
  1.42_{-0.27}^{+0.31} \times 10^{-6} &  I \\
  2.52_{-0.50}^{+0.55} \times 10^{-6} &  II\\
  \end{array}
\right.
\end{equation}
where $I(II)$ stands for the result for $I(II)$ kind of $D^{(*)}$
wave function. From the $B\to K$ transition form factor
$f_+^K(0)$, we can limit the appropriate extent of $\omega_b$.
$f_+^K(0)$ calculated from PQCD at $m_{0K}= 1.6$ GeV is consistent
with $f_+^K(0)$ by QCD sum rules \cite{18ps}, when
\begin{equation}
 0.35 \mbox{ GeV} \leq \omega_b \leq 0.46 \mbox{ GeV}.
\label{eq:wb}
\end{equation}
In the above range, the branching ratios are
\begin{equation}
 \mathrm{Br}(B^+ \to D^+ K^0) = \left \{
  \begin{array}{ll}
  0.93_{-0.10}^{+0.09} \times 10^{-6} &  I \\
  1.47_{-0.15}^{+0.15} \times 10^{-6} &  II\\
  \end{array}
\right.
\label{wb1}
\end{equation}

\begin{equation}
 \mathrm{Br}(B^+ \to D^{*+} K^0) = \left \{
  \begin{array}{ll}
  1.42_{-0.00}^{+0.00} \times 10^{-6} &  I \\
  2.52_{-0.01}^{+0.00} \times 10^{-6} &  II\\
  \end{array}
\right. \label{wb2}
\end{equation}

\begin{equation}
 \mathrm{Br}(B^+ \to D^+ K^{*0}) = \left \{
  \begin{array}{ll}
  0.96_{-0.09}^{+0.10} \times 10^{-6} &  I \\
 1.64_{-0.17}^{+0.16} \times 10^{-6} &  II\\
  \end{array}
\right. \label{wb3}
\end{equation}

\begin{equation}
 \mathrm{Br}(B^+ \to D^{*+} K^{*0}) = \left \{
  \begin{array}{ll}
  4.09_{-0.06}^{+0.06} \times 10^{-6} &  I \\
  7.04_{-0.06}^{+0.07} \times 10^{-6} &  II .\\
  \end{array}
\right. \label{wb4}
\end{equation}

\begin{table}[htbp]
\begin{center}
\begin{tabular}[t]{r|cccc}
 $M_{0K}$ & $\mathrm{Br}(B \to DK)$ & $\mathrm{Br}(B \to D^*K)$ & $\mathrm{Br}(B \to DK^*)$ & $\mathrm{Br}(B \to D^*K^*)$ \\
 \hline
 $1.12$ & $0.79$ & $0.81$ &  $-$  & $-$ \\
 $1.60$ & $0.93$ & $1.42$ &  $-$  & $-$ \\
 $2.08$ & $1.09$ & $2.22$ &  $-$  & $-$  \\
 \hline
 \hline
 $\omega_b$ & $\mathrm{Br}(B \to DK)$ &
$\mathrm{Br}(B \to D^*K)$ & $\mathrm{Br}(B \to DK^*)$ & $\mathrm{Br}(B \to D^*K^*)$\\
 \hline
 $0.28$ & $1.16$ & $1.41$ & $1.21$ & $4.25$ \\
 $0.40$ & $0.93$ & $1.42$  & $0.96$ & $4.09$ \\
 $0.52$ & $0.75$ & $1.42$ & $0.79$ & $3.99$ \\
 \hline
 \hline
 $a_{D^{(*)}}$ & $\mathrm{Br}(B \to DK)$ &
$\mathrm{Br}(B \to D^*K)$ & $\mathrm{Br}(B \to DK^*)$ & $\mathrm{Br}(B \to D^*K^*)$ \\
 \hline
 $0.21$ & $0.88$ & $1.32$ & $0.90$ & $3.79$ \\
 $0.30$ & $0.93$ & $1.42$ & $0.96$ & $4.09$ \\
 $0.39$ & $0.99$ & $1.54$ & $1.03$ & $4.40$    \\
 \end{tabular}
\end{center}
\caption{The sensitivity of the branching ratio $(10^{-6})$ to
$30$\% extent of parameters in terms of the four modes of the $B^+
\to D^{(*)+}K^{(*)0}$ decays according to the $D^{(*)}$ wave
function I.}
 \label{tb:I}
\end{table}

\begin{table}[htbp]
\begin{center}
\begin{tabular}[t]{r|cccc}
  $M_{0K}$ & $\mathrm{Br}(B \to DK)$ &
$\mathrm{Br}(B \to D^*K)$ & $\mathrm{Br}(B \to DK^*)$ & $\mathrm{Br}(B \to D^*K^*)$ \\
 \hline
 $1.12$ & $1.17$ & $1.41$ & $-$  & $-$  \\
 $1.60$ & $1.47$ & $2.52$ & $-$  & $-$  \\
 $2.08$ & $1.80$ & $3.95$ & $-$  & $-$  \\
 \hline
 \hline
 $\omega_b$ & $\mathrm{Br}(B \to DK)$ &
$\mathrm{Br}(B \to D^*K)$ & $\mathrm{Br}(B \to DK^*)$ & $\mathrm{Br}(B \to D^*K^*)$\\
 \hline
 $0.28$ & $1.86$ & $2.48$ & $2.06$ & $7.22$ \\
 $0.40$ & $1.47$ & $2.52$ & $1.64$ & $7.04$ \\
 $0.52$ & $1.20$ & $2.51$ & $1.34$ & $6.92$ \\
 \hline
 \hline
$C_D$ & $\mathrm{Br}(B \to DK)$ &
$\mathrm{Br}(B \to D^*K)$ & $\mathrm{Br}(B \to DK^*)$ & $\mathrm{Br}(B \to D^*K^*)$ \\
 \hline
 $0.56$ & $1.26$ & $-$ & $1.38$ & $-$ \\
 $0.80$ & $1.47$ & $-$ & $1.64$ & $-$ \\
 $1.04$ & $1.70$ & $-$ & $1.92$ & $-$ \\
 \hline
 \hline
  $C_{D^*}$ & $\mathrm{Br}(B \to DK)$ &
$\mathrm{Br}(B \to D^*K)$ & $\mathrm{Br}(B \to DK^*)$ & $\mathrm{Br}(B \to D^*K^*)$ \\
 \hline
 $0.49$ & $-$ & $2.13$ & $-$ & $5.99$ \\
 $0.70$ & $-$ & $2.52$ & $-$ & $7.04$ \\
 $0.91$ & $-$ & $2.95$ & $-$ & $7.50$ \\
  \end{tabular}
\end{center}
\caption{The sensitivity of the branching ratio $(10^{-6})$ to
$30$\% extent of parameters in terms of the four modes of the $B^+
\to D^{(*)+}K^{(*)0}$ decays according to the $D^{(*)}$ wave
function II.}
 \label{tb:II}
\end{table}

Besides the Perturbative annihilation contribution above, there is
also contribution from the final state interaction (FSI) in
hadronic level, such as $B^+ \to D^{(*)0}K^{(*)+}$ then
$D^{(*)0}K^{(*)+} \to D^{(*)+}K^{(*)0}$. Based on the argument of
color transparency \cite{9lb80,21bj}, FSI effects may not be
important in the two-body $B$ decays. So we suppose that the
dominant contribution is what we calculated above. The hypothesis
is consistent with the argument in \cite{6kls,22cl}.

\section{Conclusion}
In this paper, we study the four modes of $B^+ \to
D^{(*)+}K^{(*)0}$ decays. Based on the consistent PQCD framework,
we predict the branching ratios of these pure annihilation type
decays of the order of ${10}^{-6}$, and show the theoretical
errors. Such results can be measured in the two B factories in the
future.

\section*{Acknowledgments}
We thank Y. Li for the beneficial discussions. This work is partly
supported by National Science Foundation of China with contract
No.~90103013 and 10135060.

\appendix
\begin{appendix}
\section{Appendix}
\subsection{The (non-)factorizable amplitude}
At first order of $\alpha_s$, we get the analytic formulae of the
(non-)factorizable amplitude for each mode or helicity state
listed below. We neglect the small term $x_1$ in the numerators of
the hard part of $M$, since the $B$ meson wave function in eq.
(\ref{eq:Bwave}) have a sharp peak at the small $x$ region,
$\cal{O}$$(\bar \Lambda/M_B)$, where $\bar \Lambda \equiv
M_B-M_b$. It should be noticed that we do not employ this
approximation to the denominators of the propagator which are
sensitive to $x_1$. Because $x_1$ there behaves as a cut-off.
\subsubsection{$B^+ \to D^{+}K^{0}$ decay}
The amplitude for the factorizable annihilation diagrams in Fig.
1a,b is given as
\begin{eqnarray}
{F_{1}} & = & 16\pi C_F M_B^2 \int_0^1\!\!\! dx_2 dx_3
\int_0^\infty\!\!\!\!\!  b_2 db_2\, b_3 db_3\ \phi_{D}(x_2,b_2)\nonumber \\
 &\times & \Bigl[ \bigl\{ (x_3-2x_3r_2^2-r_2^2) \phi_K^A(x_3,b_3)
+ r_2 r_K \left( 1+2x_3 \right) \phi_K^P(x_3,b_3) - r_2r_K
(1-2x_3) \phi_K^T(x_3,b_3) \bigr\} \\\nonumber & &E_{f}(t_a^1)
h_a(x_2,x_3,b_2,b_3) + \bigl\{ (r_2^2-1)x_2\phi_K^A(x_3,b_3) -
2r_2 r_K(1+x_2) \phi_K^P (x_3,b_3) \bigr\} E_{f}(t_a^2)
h_a(x_3,x_2,b_3,b_2) \Bigr]. \label{eq:Fa1}
\end{eqnarray}
The amplitude for the non-factorizable annihilation diagrams in
Fig. 1a,b is obtained as
\begin{eqnarray}
M_{1} & = & -\frac{1}{\sqrt{2N_c}} 64\pi C_F M_B^2 \int_0^1\!\!\!
dx_1 dx_2 dx_3
 \int_0^\infty\!\!\!\!\! b_1 db_1\, b_2 db_2\
 \phi_B(x_1,b_1) \phi_{D}(x_2,b_2)\nonumber \\
&\times &\Bigl[ \bigl\{ (x_3+(x_2-2 x_3-1)r_2^2) \phi_K^A(x_3,b_2)
 + r_2  r_K \left( 2+x_2+x_3\right) \phi_K^P(x_3,b_2) - r_2 r_K \left(x_2-x_3\right) \phi_K^T(x_3,b_2)
\bigr\}\nonumber \\
 & &E_{m}(t_{m}^1) h_a^{(1)}(x_1, x_2,x_3,b_1,b_2)\nonumber \\
& &-\bigl\{
 x_2 \phi_K^A(x_3,b_2)
 + r_2 r_K \left(x_2+x_3 \right)  \phi_K^P(x_3,b_2) + r_2 r_K \left(x_2-x_3 \right)  \phi_K^T(x_3,b_2)
\bigr\}E_{m}(t_{m}^2) h_a^{(2)}(x_1, x_2,x_3,b_1,b_2) \Bigr].
\label{eq:Ma1}
\end{eqnarray}
where $C_F = 4/3$ is the group factor of $\mathrm{SU}(3)_c$ gauge
group, and $r_K = M_{0K}/M_B$, and the functions $E_{f}$, $E_{m}$,
$t_a^{1,2}$, $h_a$ are given in the appendix A. 3.
\subsubsection{$B^+ \to D^{*+}K^{0}$ decay}
\begin{eqnarray}
F_{2} & = & -16\pi C_F M_B^2 \int_0^1\!\!\! dx_2 dx_3
 \int_0^\infty\!\!\!\!\!  b_2 db_2\, b_3 db_3\ \phi_{D^*}(x_2,b_2) \nonumber \\
&  \times & \Bigl[ \bigl\{ (x_3-2x_3r_2^2+r_2^2) \phi_K^A(x_3,b_3)
+ r_2 r_K   \phi_K^P(x_3,b_3)  - r_2 r_K \phi_K^T(x_3,b_3)
\bigr\} E_{f}(t_a^1) h_a(x_2,x_3,b_2,b_3) \nonumber \\
& & - \bigl\{ (1-r_2^2)x_2\phi_K^A(x_3,b_3) -2r_2 r_K(1-x_2)
\phi_K^P (x_3,b_3) \bigr\} E_{f}(t_a^2) h_a(x_3,x_2,b_3,b_2)
\Bigr], \label{eq:Fa2}
\end{eqnarray}
\begin{eqnarray}
M_{2}& =&  \frac{1}{\sqrt{2N_c}} 64\pi C_F M_B^2 \int_0^1\!\!\!
dx_1 dx_2 dx_3
 \int_0^\infty\!\!\!\!\! b_1 db_1\, b_2 db_2\
 \phi_B(x_1,b_1) \phi_{D^*}(x_2,b_2) \nonumber \\
&  \times & \Bigl[ \bigl\{ (x_3+(1-x_2-2 x_3)r_2^2)
\phi_K^A(x_3,b_2)
 + r_2  r_K \left( x_3-x_2\right) \phi_K^P(x_3,b_2) \nonumber \\
& &
 - r_2 r_K \left(2-x_2-x_3\right) \phi_K^T(x_3,b_2)
\bigr\}
E_{m}(t_{m}^1) h_a^{(1)}(x_1, x_2,x_3,b_1,b_2) \nonumber \\
& & - \bigl\{
 (1-2r_2^2)x_2 \phi_K^A(x_3,b_2)
 + r_2 r_K \left(x_2-x_3 \right)  \phi_K^P(x_3,b_2) \nonumber \\
& &
 + r_2 r_K \left(x_2+x_3 \right)  \phi_K^T(x_3,b_2)
\bigr\}E_{m}(t_{m}^2) h_a^{(2)}(x_1, x_2,x_3,b_1,b_2) \Bigr].
\label{eq:Ma2}
\end{eqnarray}
\subsubsection{$B^+ \to D^{+}K^{*0}$ decay}
\begin{eqnarray}
F_{3} & =& -16\pi C_F M_B^2 \int_0^1\!\!\! dx_2 dx_3
 \int_0^\infty\!\!\!\!\!  b_2 db_2\, b_3 db_3\ \phi_{D}(x_2,b_2) \nonumber \\
& \times & \Bigl[ \bigl\{ (x_3-2x_3r_2^2-r_2^2)
\phi_{K^*}(x_3,b_3) + r_2 r_3 \left( 1+2x_3\right)
\phi_{K^*}^s(x_3,b_3)  - r_2 r_3 (1-2x_3) \phi_{K^*}^t(x_3,b_3)
\bigr\}\nonumber \\
& & E_{f}(t_a^1) h_a(x_2,x_3,b_2,b_3) \nonumber \\
& & - \bigl\{ (1-r_2^2)x_2\phi_{K^*}(x_3,b_3) +2r_2 r_3(1+x_2)
\phi_{K^*}^s (x_3,b_3) \bigr\}E_{f}(t_a^2) h_a(x_3,x_2,b_3,b_2)
\Bigr], \label{eq:Fa3}
\end{eqnarray}
\begin{eqnarray}
M_{3} & =&  \frac{1}{\sqrt{2N_c}} 64\pi C_F M_B^2 \int_0^1\!\!\!
dx_1 dx_2 dx_3
 \int_0^\infty\!\!\!\!\! b_1 db_1\, b_2 db_2\
 \phi_B(x_1,b_1) \phi_{D}(x_2,b_2) \nonumber \\
& & \times \Bigl[ \bigl\{ (x_3+(x_2-2 x_3-1)r_2^2)
\phi_{K^*}(x_3,b_2)
 + r_2  r_3 \left( 2+x_2+x_3\right) \phi_{K^*}^s(x_3,b_2) \nonumber \\
& &
 - r_2 r_3 \left(x_2-x_3\right) \phi_{K^*}^t(x_3,b_2)
\bigr\}
E_{m}(t_{m}^1) h_a^{(1)}(x_1, x_2,x_3,b_1,b_2) \nonumber \\
& & - \bigl\{ x_2 \phi_{K^*}(x_3,b_2)
 + r_2 r_3 \left(x_2+x_3 \right)  \phi_{K^*}^s(x_3,b_2) \nonumber \\
& &
 + r_2 r_3 \left(x_2-x_3 \right)  \phi_{K^*}^t(x_3,b_2)
\bigr\}E_{m}(t_{m}^2) h_a^{(2)}(x_1, x_2,x_3,b_1,b_2) \Bigr].
\label{eq:Ma3}
\end{eqnarray}

\subsubsection{$B^+ \to D^{*+}K^{*0}$ decay}
\begin{eqnarray}
F_{L} & =& 16\pi C_F M_B^2 \int_0^1\!\!\! dx_2 dx_3
 \int_0^\infty\!\!\!\!\!  b_2 db_2\, b_3 db_3\ \phi_{D^*}(x_2,b_2) \nonumber \\
& & \times \Bigl[ \bigl\{ (x_3-2x_3r_2^2+r_2^2)
\phi_{K^*}(x_3,b_3) + r_2 r_3   \phi_{K^*}^s(x_3,b_3) - r_2 r_3
\phi_{K^*}^t(x_3,b_3)
\bigr\} E_{f}(t_a^1) h_a(x_2,x_3,b_2,b_3) \nonumber \\
& & - \bigl\{ (1-r_2^2)x_2\phi_{K^*}(x_3,b_3) +2r_2 r_3(x_2-1)
\phi_{K^*}^s (x_3,b_3) \bigr\} E_{f}(t_a^2) h_a(x_3,x_2,b_3,b_2)
\Bigr], \label{eq:FaL4}
\end{eqnarray}
\begin{eqnarray}
M_{L} &= & \frac{1}{\sqrt{2N_c}} 64\pi C_F M_B^2 \int_0^1\!\!\!
dx_1 dx_2 dx_3
 \int_0^\infty\!\!\!\!\! b_1 db_1\, b_2 db_2\
 \phi_B(x_1,b_1) \phi_{D^*}(x_2,b_2) \nonumber \\
& & \times \Bigl[ \bigl\{ (-x_3+(x_2+2 x_3-1)r_2^2)
\phi_{K^*}(x_3,b_2)
 + r_2  r_3 \left( x_2-x_3\right) \phi_{K^*}^s(x_3,b_2) \nonumber \\
& &
 + r_2 r_3 \left(2-x_2-x_3\right) \phi_{K^*}^t(x_3,b_2)
\bigr\}
E_{m}(t_{m}^1) h_a^{(1)}(x_1, x_2,x_3,b_1,b_2) \nonumber \\
& & + \bigl\{
 \left(1-2r_2^2 \right) x_2
 \phi_{K^*}(x_3,b_2)
 + r_2 r_3 \left(x_2-x_3 \right)  \phi_{K^*}^s(x_3,b_2) \nonumber \\
& &
 + r_2 r_3 \left(x_2+x_3\right)  \phi_{K^*}^t(x_3,b_2)
\bigr\} E_{m}(t_{m}^2) h_a^{(2)}(x_1, x_2,x_3,b_1,b_2) \Bigr].
\label{eq:MaL4}
\end{eqnarray}

\begin{eqnarray}
F_{T1} &=& 16\pi C_F M_B^2 \int_0^1\!\!\! dx_2 dx_3
 \int_0^\infty\!\!\!\!\!  b_2 db_2\, b_3 db_3\ \phi_{D^*}(x_2,b_2) \nonumber \\
& & \times \Bigl[ \bigl\{- 2r_2 r_3x_3 \phi_{K^*}^v(x_3,b_3) -2
r_2 r_3 x_3 \phi_{K^*}^a(x_3,b_3) +2 r_2^2 \phi_{K^*}^T(x_3,b_3)
\bigr\} E_{f}(t_a^1) h_a(x_2,x_3,b_2,b_3) \nonumber \\
& & + \bigl\{ 2 r_2 r_3 \phi_{K^*}^v(x_3,b_3) +2r_2 r_3
\phi_{K^*}^a (x_3,b_3) \bigr\} E_{f}(t_a^2) h_a(x_3,x_2,b_3,b_2)
\Bigr], \label{eq:FaT14}
\end{eqnarray}
\begin{eqnarray}
M_{T1} &=&  \frac{1}{\sqrt{2N_c}} 64\pi C_F M_B^2 \int_0^1\!\!\!
dx_1 dx_2 dx_3
 \int_0^\infty\!\!\!\!\! b_1 db_1\, b_2 db_2\
 \phi_B(x_1,b_1) \phi_{D^*}(x_2,b_2) \nonumber \\
&  \times & \Bigl[ \bigl\{ 2r_2  r_3 \phi_{K^*}^v(x_3,b_2)
 +2 r_2  r_3  \phi_{K^*}^a(x_3,b_2)
 -2 r_2^2 \left(1-x_2\right) \phi_{K^*}^T(x_3,b_2)
\bigr\}\nonumber \\
& & E_{m}(t_{m}^1) h_a^{(1)}(x_1, x_2,x_3,b_1,b_2) - \bigl\{
2r_2^2 x_2 \phi_{K^*}^T(x_3,b_2) \bigr\} E_{m}(t_{m}^2)
h_a^{(2)}(x_1, x_2,x_3,b_1,b_2) \Bigr]. \label{eq:MaT14}
\end{eqnarray}

\begin{eqnarray}
F_{T2} &=& 16\pi C_F M_B^2 \int_0^1\!\!\! dx_2 dx_3
 \int_0^\infty\!\!\!\!\!  b_2 db_2\, b_3 db_3\ \phi_{D^*}(x_2,b_2) \nonumber \\
&  \times & \Bigl[ \bigl\{ -2r_2 r_3\phi_{K^*}^v(x_3,b_3) +2 r_2
r_3
\phi_{K^*}^a(x_3,b_3) \bigr\} E_{f}(t_a^1) h_a(x_2,x_3,b_2,b_3) \nonumber \\
& & + \bigl\{ 2 r_2 r_3x_2 \phi_{K^*}^v(x_3,b_3) -2r_2 r_3 x_2
\phi_{K^*}^a (x_3,b_3) \bigr\} E_{f}(t_a^2) h_a(x_3,x_2,b_3,b_2)
\Bigr], \label{eq:FaT24}
\end{eqnarray}
\begin{eqnarray}
M_{T2} &= & \frac{1}{\sqrt{2N_c}} 64\pi C_F M_B^2 \int_0^1\!\!\!
dx_1 dx_2 dx_3
 \int_0^\infty\!\!\!\!\! b_1 db_1\, b_2 db_2\
 \phi_B(x_1,b_1) \phi_{D^*}(x_2,b_2) \nonumber \\
& \times &  \bigl\{ 2r_2  r_3 \phi_{K^*}^v(x_3,b_2) -2 r_2  r_3
\phi_{K^*}^a(x_3,b_2)
 \bigr\}E_{m}(t_{m}^1) h_a^{(1)}(x_1, x_2,x_3,b_1,b_2). \label{eq:MaT24}
\end{eqnarray}
where the subscript $L(T(1,2))$ , i.e. the helicity states of the
two vector mesons $\sigma$ in eq. (\ref{dr1}), stands for the
longitudinal (transverse) component respectively. Conveniently we
choose the polarization state $T1$ as $\epsilon_{2T}=
\frac{1}{\sqrt{2}}(0,0,1, i) , \epsilon_{3T}=
\frac{1}{\sqrt{2}}(0,0,1,- i)$, $T2$ as $\epsilon_{2T}=
\frac{1}{\sqrt{2}}(0,0,1, -i) , \epsilon_{3T}=
\frac{1}{\sqrt{2}}(0,0,1,i)$. Each amplitude $A_\sigma$ also is
the sum of two parts, factorizable and non-factorizable diagrams,
related by eq. (\ref{eq:neut_amp}).

\subsection{The $K^{(*)}$ meson wave functions}
The $K$ and $K^*$ meson wave functions are given
as\cite{18ps,19vect}
\begin{eqnarray}
\phi_K^A(x) &=& \frac{f_K}{2\sqrt{2 N_c}} 6 x(1-x) \left\{ 1 +0.51
(1-2x) +
0.2  C^{3/2}_2(1-2x) \right\}, \\
\phi_K^P(x) &=& \frac{f_K}{2\sqrt{2 N_c}} \left\{ 1 + 0.212
C^{1/2}_2(1-2x)
-0.148  C^{1/2}_4(1-2x) \right\}, \\
\phi_K^T(x) &=& \frac{f_K}{2\sqrt{2 N_c}} (1-2x) \left\{ 1 +
0.1581[-3 + 5 (1-2x)^2] \right\},
\end{eqnarray}

\begin{eqnarray}
\phi_{K^* }( x) &=&\frac{f_{K^*}}{2\sqrt{2N_{c}}}6x(1-x)
\Big[1+0.57(1-2x)+0.07C^{3/2}_2(1-2x)\Big]\;,
\label{pk2}\\
\phi_{K^*}^{t}( x) &=&\frac{f_{K^*}^T}{2\sqrt{2N_{c}}} \bigg\{
0.3(1-2x)\left[3(1-2x)^2+10(1-2x)-1\right]+1.68C^{1/2}_4(1-2x)
\nonumber \\
&&+0.06(1-2x)^2\left[5(1-2x)^2-3\right] +0.36\left[
1-2(1-2x)(1+\ln(1-x))\right] \bigg\} \;,
\label{pk3t}\\
\phi _{K^*}^s( x)  &=&\frac{f_{K^*}^T}{2\sqrt{2N_{c}}} \bigg\{
3(1-2x)\left[1+0.2(1-2x)+0.6(10x^2-10x+1)\right]
\nonumber \\
& &-0.12x(1-x)+0.36[1-6x-2\ln(1-x)]\bigg\} \;, \label{pk3s}
\end{eqnarray}

\begin{eqnarray}
\phi_{K^*}^{T}(x)&=&\frac{f_{K^*}^T}{2\sqrt{2N_c}} 6x(1-x)
\Big[1+0.6(1-2x)+0.04C^{3/2}_2(1-2x)\Big]\;,
\label{pkt}\\
\phi_{K^*}^v(x)&=&\frac{f_{K^*}}{2\sqrt{2N_c}}
\bigg\{\frac{3}{4}\Big[1+(1-2x)^2+0.44(1-2x)^3\Big]
+0.4C^{1/2}_2(1-2x)
\nonumber \\
&& +0.88C^{1/2}_4(1-2x)+0.48[2x+\ln(1-x)] \bigg\}\;,
\label{pkv}\\
\phi_{K^*}^a(x) &=&\frac{f_{K^*}}{4\sqrt{2N_{c}}}
\Big\{3(1-2x)\Big[1+0.19(1-2x)+0.81(10x^2-10x+1)\Big]\nonumber \\
&&-1.14x(1-x)+0.48[1-6x-2\ln(1-x)]\Big\}\;, \label{pka}
\end{eqnarray}
with the Gegenbauer polynomials,
\begin{eqnarray}
C_2^{1/2}(\xi)=\frac{1}{2}(3\xi^2-1)\;,\;\;\;\;
C_4^{1/2}(\xi)=\frac{1}{8}(35 \xi^4 -30 \xi^2 +3)\;,\;\;\;\;
C_2^{3/2}(\xi)=\frac{3}{2}(5\xi^2-1)\;.
\end{eqnarray}

\subsection{Some used formulae}
The definitions of some functions used in the text are presented
in this appendix. In the numerical analysis we use one loop
expression for strong coupling constant,
\begin{equation}
 \alpha_s (\mu) = \frac{4 \pi}{\beta_0 \ln (\mu^2 / \Lambda^2)},
\label{eq:alphas}
\end{equation}
where $\beta_0 = (33-2n_f)/3$ and $n_f$ is number of active flavor
at appropriate scale. $\Lambda$ is QCD scale, which we use as
$250$ MeV at $n_f=4$. We also use leading logarithms expressions
for Wilson coefficients $C_{1,2}$ presented in
ref.~\cite{14buras}.

The function $E_f$ and $E_m$ including Wilson coefficients are
defined as
\begin{eqnarray}
 & &E_{f}(t) = a(t) \alpha_s(t)\, e^{-S_D(t)-S_K(t)}, \\
 & & E_{m}(t) = C_1(t) \alpha_s(t)\, e^{-S_B(t)-S_D(t)-S_K(t)}|_{b_3=b_2},
\end{eqnarray}
 where
\begin{equation}
 a(t) =  C_2(t)+ \frac{C_1(t)}{N_c},\quad
\end{equation}
and $S_B$, $S_D$, and $S_K$ result from summing both double
logarithms caused by soft gluon corrections and single ones due to
the renormalization of ultra-violet divergence. The above $S_{B,
D, K}$ are defined as
\begin{eqnarray}
& &S_B(t) = s(x_1P_1^+,b_1) +
2 \int_{1/b_1}^t \frac{d\mu'}{\mu'} \gamma_q(\mu'), \\
& & S_D(t) = s(x_2P_2^+,b_2) +
2 \int_{1/b_2}^t \frac{d\mu'}{\mu'} \gamma_q(\mu'), \\
& &S_K(t) = s(x_3P_3^-,b_3) + s((1-x_3)P_3^-,b_3) + 2
\int_{1/b_3}^t \frac{d\mu'}{\mu'} \gamma_q(\mu'),
\end{eqnarray}
where $s(Q,b)$, so-called Sudakov factor, is given as \cite{23lm}
\begin{eqnarray}
  s(Q,b) &=& \int_{1/b}^Q \!\! \frac{d\mu'}{\mu'} \left[
 \left\{ \frac{2}{3}(2 \gamma_E - 1 - \ln 2) + C_F \ln \frac{Q}{\mu'}
 \right\} \frac{\alpha_s(\mu')}{\pi} \right. \nonumber \\
& &  \left.+ \left\{ \frac{67}{9} - \frac{\pi^2}{3} -
\frac{10}{27} n_f
 + \frac{2}{3} \beta_0 \ln \frac{e^{\gamma_E}}{2} \right\}
 \left( \frac{\alpha_s(\mu')}{\pi} \right)^2 \ln \frac{Q}{\mu'}
 \right],
 \label{eq:SudakovExpress}
\end{eqnarray}
 $\gamma_E=0.57722\cdots$ is Euler constant,
and $\gamma_q = -\alpha_s/\pi$ is the quark anomalous dimension.

The functions $h_a$, $h_a^{(1)}$, and $h_a^{(2)}$  in the decay
amplitudes consist of two parts: one is the jet function
$S_t(x_i)$ derived by the threshold resummation\cite{13sum}, the
other is the propagator of virtual quark and gluon. They are
defined by
\begin{eqnarray}
 h_a(x_2,x_3,b_2,b_3)& = &S_t(1-x_3)\left( \frac{\pi i}{2}\right)^2
H_0^{(1)}(M_B\sqrt{(1-r_2^2)x_2x_3}\, b_2) \nonumber \\
&\times &\left\{ H_0^{(1)}(M_B\sqrt{(1-r_2^2)x_3}\, b_2)
J_0(M_B\sqrt{(1-r_2^2)x_3}\, b_3) \theta(b_2 - b_3) + (b_2
\leftrightarrow b_3 ) \right\},
\label{eq:propagator1} \\
\end{eqnarray}
\begin{eqnarray}
h^{(j)}_a(x_1,x_2,x_3,b_1,b_2)  =&  \biggl\{ \frac{\pi i}{2}
\mathrm{H}_0^{(1)}(M_B\sqrt{(1-r_2^2)x_2x_3}\, b_1)
 \mathrm{J}_0(M_B\sqrt{(1-r_2^2)x_2x_3}\, b_2) \theta(b_1-b_2)+ (b_1 \leftrightarrow b_2) \biggr\}&
\nonumber \\
& \times\left(
\begin{array}{cc}
 {\mathrm{K}}_0(M_B F_{(j)} b_1), & \text{for}\quad F^2_{(j)}>0 \\
 \frac{\pi i}{2} \mathrm{H}_0^{(1)}(M_B\sqrt{|F^2_{(j)}|}\ b_1), &
 \text{for}\quad F^2_{(j)}<0
\end{array}\right),&
\label{eq:propagator_3}
\end{eqnarray}
where $\mathrm{H}_0^{(1)}(z) = \mathrm{J}_0(z) + i\,
\mathrm{Y}_0(z)$, and $F_{(j)}$s are defined by
\begin{eqnarray}
 &&F^2_{(1)} =  x_1+x_2+(1-x_1- x_2)x_3(1-r_2^2),\\
 &&F^2_{(2)} = x_3(x_1-x_2)(1-r_2^2).
\end{eqnarray}

We adopt the parametrization for $S_t(x)$ of the factorizable
contributions,
\begin{equation}
 S_t(x) = \frac{2^{1+2c}\Gamma(3/2 +c)}{\sqrt{\pi} \Gamma(1+c)}
[x(1-x)]^c,\quad c = 0.3.
\end{equation}
In the non-factorizable annihilation contributions, $S_t(x)$ gives
a very small numerical effect to the amplitude\cite{13sum}.
Therefore, we drop $S_t(x)$ in $h_a^{(1)}$ and $h_a^{(2)}$. The
hard scale $t$'s in the amplitudes are taken as the largest energy
scale in the $H$ to kill the large logarithmic radiative
corrections:
\begin{eqnarray}
&& t_a^1 = \mathrm{max}(M_B \sqrt{(1-r_2^2)x_3},1/b_2,1/b_3), \\
&& t_a^2 = \mathrm{max}(M_B \sqrt{(1-r_2^2)x_2},1/b_2,1/b_3), \\
&& t_{m}^j = \mathrm{max}(M_B \sqrt{|F^2_{(j)}|}, M_B
\sqrt{((1-r_2^2)x_2x_3 }, 1/b_1,1/b_2).
\end{eqnarray}
\end{appendix}



\begin{thebibliography}{99}

\bibitem{1wsb} M. Wirbel, B. Stech, M. Bauer, Z. Phys. C29, 637 (1985);
 M. Bauer, B. Stech, M. Wirbel, Z. Phys. C34, 103 (1987);
L.-L. Chau, H.-Y. Cheng, W.K. Sze, H. Yao, B. Tseng, Phys. Rev.
D43, 2176 (1991), Erratum: D58, 019902 (1998).

\bibitem{2aklch} A. Ali, G. Kramer and C.D. L\"u, Phys. Rev. D58, 094009
(1998); C.D. L\"u, Nucl. Phys. Proc. Suppl. 74, 227-230 (1999);
Y.-H. Chen, H.-Y. Cheng, B. Tseng, K.-C. Yang, Phys. Rev. D60,
094014 (1999); H.-Y. Cheng and K.-C. Yang, Phys. Rev. D62, 054029
(2000).

\bibitem{3cly} H.Y. Cheng, H-n. Li, and K.C. Yang, Phys. Rev. D 60,
094005 (1999).

\bibitem{4yl} T.W. Yeh and H-n. Li, Phys. Rev. D 56, 1615 (1997).

\bibitem{5bbns} M. Beneke, G. Buchalla, M. Neubert, C.T. Sachrajda,
Phys. Rev. Lett. 83, 1914 (1999); Nucl. Phys. B591, 313 (2000).

\bibitem{6kls} Y.-Y. Keum, H.-n. Li and A. I. Sanda,
Phys. Lett. B504, 6 (2001); Phys. Rev. D63, 054008 (2001).

\bibitem{7luy} C.-D. L\"u, K. Ukai and M.-Z. Yang,
Phys. Rev. D63, 074009 (2001); C.-D. L\"u,   pp.  173-184,
Proceedings of International Conference on Flavor Physics (ICFP
2001), World Scientific, 2001, hep-ph/0110327.

\bibitem{8lb79} G.P. Lepage and S.J. Brodsky, Phys. Lett. B 87,
359 (1979).

\bibitem{9lb80} G.P. Lepage and S.J. Brodsky, Phys. Rev. D 22, 2157
(1980).


\bibitem{10chang} C.-H. V. Chang and H.-n. Li, Phys. Rev. D55, 5577 (1997);
T.-W. Yeh and H.-n. Li, Phys. Rev. D56, 1615 (1997).

\bibitem{11PQCD} H.-n. Li, Phys. Rev. D64, 014019 (2001); S. Mishima, Phys. Lett.
B521, 252 (2001); E. Kou and A.I. Sanda, Phys. Lett. B525, 240
(2002); C.-H. Chen, Y.-Y. Keum, and H.-n. Li, Phys. Rev. D64,
112002 (2001); C.-D. L\"u and M.Z. Yang, Eur. Phys. J. C23, 275
(2002); A.I. Sanda and K. Ukai, Prog. Theor. Phys. 107, 421
(2002);  C.-H. Chen, Y.-Y. Keum, and H.-n. Li, Phys. Rev. D66,
054013 (2002); M. Nagashima and H.-n. Li, hep-ph/0202127; Y.-Y.
Keum, hep-ph/0209002; hep-ph/0209208(to appear in PRL);
hep-ph/0210127; Y.-Y. Keum and A. I. Sanda, Phys.Rev. D 67, 054009
(2003).

\bibitem{112} C.D. L\"u, Eur. Phys. J. C24, 121 (2002); Y. Li, C.D.
L\"u, J.Phys. G29 2115 (2003); hep-ph/0305278; hep-ph/0308243;
C.D. L\"u, K. Ukai, Eur.Phys.J. C28 305 (2003).

\bibitem{12soft} H.-n. Li and B. Tseng, Phys. Rev. D57, 443 (1998).

\bibitem{13sum} H.-n. Li, Phys. Rev. D 66, 094010 (2002); H.-n. Li, K. Ukai, Phys. Lett. B 555, 197 (2003).

\bibitem{14buras} G. Buchalla, A. J. Buras and
    M. E. Lautenbacher, Rev. Mod. Phys. 68, 1125(1996).

\bibitem{15pdgv} Review of Particle Physics, K. Hagiwara {\it et al.}, Phys. Rev. D66, 010001 (2002).

\bibitem{16kls} T. Kurimoto, H.-n. Li, and A. I. Sanda,
Phys. Rev. D 67, 054028 (2003).

\bibitem{17many} Y.-Y. Keum, T. Kurimoto, H.-N. Li, C.-D. L\"u, A.I. Sanda, hep-ph/0305335.

\bibitem{18ps} P. Ball, JHEP, 09, 005, (1998); JHEP, 01, 010, (1999).


\bibitem{19vect} P. Ball, V.M. Braun, Y. Koike, and K. Tanaka, Nucl. Phys.
B 529, 323 (1998); P. Ball, V. M. Braun, hep-ph/9808229.

\bibitem{20pdg} Particle Data Group, Phys. Rev. D66, Part I (2002).

\bibitem{21bj} J.D. Bjorken, Nucl. Phys. B (Proc. Suppl.) 11, 325 (1989).


\bibitem{22cl} C.-H. Chen and H.-n. Li, Phys. Rev. D63, 014003 (2001).

\bibitem{23lm} H.-n. Li, B. Melic, Eur. Phys. J. C11, 695
(1999).



   \end{thebibliography}
\end{document}